# Charge transfer mediated giant photo-amplification in air-stable α-CsPbI$_3$ nanocrystals decorated 2D-WS$_2$ photo-FET with asymmetric contacts

Shreyasi Das[1], Arup Ghorai[1,2], Sourabh Pal[3], Somnath Mahato[1], Soumen Das[4], Samit K. Ray[5] *

[1]*School of Nano Science and Technology, IIT Kharagpur, Kharagpur, West Bengal, India, 721302*
[2]*Department of Materials Science and Engineering, Pohang University of Science and Technology, Pohang 790-784, Korea*
[3]*Advanced Technology Development Centre, IIT Kharagpur, Kharagpur, West Bengal, India, 721302*
[4]*School of Medical Science and Technology, IIT Kharagpur, Kharagpur, West Bengal, India, 721302*
[5]*Department of Physics, IIT Kharagpur, Kharagpur, West Bengal, India, 721302*
Email : physkr@phy.iitkgp.ac.in

## Abstract

Hybrid heterostructure based phototransistors are attractive owing to their high gain induced by photogating effect. However, the absence of an in-plane built-in electric field in the single channel layer transistor results in a relatively higher dark current and require a large operating gate voltage of the device. Here, we report novel air-stable cesium lead iodide/tungsten di-sulfide (CsPbI$_3$/WS$_2$) mixed dimensional heterostructure based photo-field-effect-transistors (photo-FETs) with asymmetric metal electrodes (Cr/WS$_2$/Au), exhibiting extremely low dark current (~$10^{-12}$ A) with a responsivity of ~ $10^2$ A/W at zero gate bias. The Schottky barrier (WS$_2$/Au interface) induced rectification characteristics in the channel accompanied by the excellent photogating effect from solution-processed α-phase CsPbI$_3$ NCs sensitizers, resulting in gate-tunable broadband photodetection with a very high responsivity (~$10^4$ A/W) and excellent sensitivity (~$10^6$). Most interestingly, the device shows superior performance even under high humidity (50-65%) conditions owing to the formation of cubic α-phase CsPbI$_3$ nanocrystals with a relatively smaller lattice constant ($a$ = 6.2315 Å) and filling of surface vacancies (Pb$^{2+}$ centres) with the sulfur atoms from WS$_2$ layer, thus protecting it from environmental degradation. These results emphasise a novel strategy for developing mixed dimensional hybrid heterostructure based phototransistors for futuristic integrated nano-optoelectronic systems.

**Keywords:** Two dimensional TMDs, Inorganic perovskites, Sensitizers, Asymmetric electrodes, Mixed dimensional phototransistors

# Introduction

Fabrication of high performance phototransistors demands superior channel material, which should be of high carrier mobility for high gain bandwidth product, a direct bandgap for efficient optical absorption, a thinner layer for full depletion leading to ultralow dark current and very low trap state density for low subthreshold swings. Layered semiconducting two dimensional (2D) TMDs fulfil most of these requirements [1–4] making them a potential candidate for photo-field-effect transistor (photo-FET). However, some trade-offs are found in single semiconductor channel based phototransistors, due to the simultaneous occurrence of both absorption and amplification processes within the same layer, compromising the device performance [5–7]. To overcome this shortcoming as well as for fabrication simplicity, attention has been paid on few-layer TMDs over monolayer with a wider spectral photoresponse and relatively higher absorbance [8–10], which in turn increases the dark current values requiring a higher gate voltage to operate the device in full depletion mode [11]. This issue can be addressed via incorporating a Schottky barrier by judiciously selecting the metal contacts and utilizing the developed depletion region to facilitate an unidirectional current transport, which leads to the significant lowering of dark current in few layered TMD based photo-FETs at zero gate bias [12–14]. The built-in electric field of the Schottky interface further helps in the separation of photogenerated charge carriers across the channel layer [15] leading to zero gate bias driven enhanced photosensitivity, although the responsivity of these devices are limited owing to the lower absorption coefficient of 2D TMD material. For the further improvement of photoresponsivity, recent works are focused on sensitizing the thin channel material with semiconductor nanocrystals (NCs), referred to as sensitizers, having excellent absorption characteristics and opposite doping polarity to create a vertical junction for subsequent charge separation [16–19].

Recently, all-inorganic cesium lead halide (CsPbX$_3$: X = I, Br, and Cl) perovskite NCs have drawn tremendous interests in the field of optoelectronics [20], featuring superior emissive (approaching photoluminescence quantum yield ~ 100%) [21] characteristics, extremely high absorption coefficient [22], large carrier diffusion lengths (>3 mm), fast radiative recombination rates and most importantly their improved stability [23] over organic-inorganic hybrid perovskites [24,25]. Especially colloidal synthesised zero-dimensional (0D) α-CsPbI$_3$ NCs exhibit extended photoabsorption band covering the whole visible spectrum with high absorption coefficient (~ $3 \times 10^4$ cm$^{-1}$ at 680 nm) [26] and better photostability, making them attractive as a photoactive material for high performance optoelectronic devices under ambient condition. Weerd et al. have recently reported colloidal CsPbI$_3$ NCs with high quantum yield of ~ 98% prepared by hot-injection method which reveals highly efficient carrier multiplication as well as longer build-up of free carrier concentration [27]. Among four existing phases (α-cubic, β-tetragonal, γ-orthorhombic, and δ-orthorhombic) of CsPbI$_3$ NCs, cubic α-phase has high stability due to its low surface-to-volume ratio (lattice constant $a$ = 6.23 Å) without any octahedral inclination or lattice distortion in the [PbI$_6$]$^{4-}$ octahedra as well as in the unit cell [28–30]. These extraordinary properties of α-phase CsPbI$_3$ NCs provoke to utilize them as an effective sensitizer in 2D channel based hybrid phototransistor devices. However, owing to the vulnerability towards moisture for halide perovskite NCs, recently, the surface passivation of CsPbI$_3$ via coordination of Pb$^{2+}$ centres with sulphur donors has been explored to protect them from environmental degradation [31] to improve the device stability under ambient conditions.

In this work, we present a proof-of-concept for 0D/2D CsPbI$_3$/WS$_2$ based mixed dimensional van der Waals heterostructure (MvWH) photo-FETs utilizing the superior photoabsorption attributes of α-phase CsPbI$_3$ NCs as sensitizers, along with a sub 5-nm thick 2D WS$_2$ channel which acts as an expressway for carrier transport. In our device configuration, the developed built-in electrical field at the heterostructure interface facilitates efficient transfer of photogenerated carriers from CsPbI$_3$ NCs into the WS$_2$ layer resulting in an excellent broadband visible photoabsorption in the hybrid system. Moreover, asymmetric metal contacts (Au-Cr) as source and drain electrodes are purposefully chosen to utilize the large built-in potential along the WS$_2$/Au Schottky junction for a diode-like unidirectional current flow and effective separation of photogenerated electron-hole pairs, leading to an excellent rectifying ratio of ~ 10$^4$ with a very low dark current of the order of ~ pA under a low source-to-drain bias ($V_{DS}$) without any applied back gate voltage. Fabricated hybrid phototransistor devices exhibit a very high photoresponsivity of ~ 10$^4$ A/W at ~ 40 V gate bias upon visible light illumination (~ 0.1 µW) due to the photogating effect and a broad spectral bandwidth across the entire visible range. In addition, the coordination of sulphur atoms of WS$_2$ layer with Pb$^{2+}$ centres of CsPbI$_3$ NCs reveals an excellent device stability under 50-65% humidity condition. Our results not only open up new avenues for studying fundamental carrier transport and relaxation pathways in hybrid van der Waals heterojunctions, but also pave the way for constructing high-performance optoelectronic devices using mixed-dimensional 0D/2D hybrid building blocks, rather than using purely 2D layered materials.

## Experimental section:

**Synthesis of α-phase CsPbI$_3$ NCs:**

This is a two-step reaction, where, in the first step Caesium oleate (Cs-OA) was prepared followed by the synthesis of α-phase CsPbI$_3$ NCs in the second step.

i) Firstly, caesium carbonate (Cs$_2$CO$_3$) of 814 mg and 40 ml Olive oil were added in a round bottom two-neck flask, which was heated at 120⁰C for 1 hr under vacuum condition. Followed by the rise in temperature to 150⁰C under N$_2$ atmosphere for 10-15 mins, the desired transparent solution of Cs-OA was stored for further use to synthesize CsPbI$_3$.

ii) Next, 870 mg of PbI$_2$ and 5 ml of olive oil (instead of 1-octadecene (ODE)) were mixed which was heated to 120⁰C under vacuum for 1hr. Thereafter, we swiftly injected 1 ml of

Oleyl amine (OLAm) in the reaction mixture under $N_2$ atmosphere to get a transparent solution. Finally, preheated (100ºC) Cs-OA was injected to the reaction mixture and cooled immediately in an ice bath to quench the reaction to obtain desired α-phase $CsPbI_3$ NCs.

Finally, as-synthesized $CsPbI_3$ was purified through centrifugation using excess hexane. The centrifugation process (20000 rpm) was repeated for several times to remove excess OLAm/olive oil from the product. Finally, the sedimentation was collected and re-dispersed in hexane. The collected dispersion was stored in a sealed vial for further characterisations and device fabrications.

**$CsPbI_3$/$WS_2$ MvWH photo-FET device fabrication:**

For the fabrication of the $CsPbI_3$/$WS_2$ MvWH photo-FET device, $WS_2$ flakes were mechanically exfoliated from the bulk $WS_2$ crystal (2D semiconductor Inc. Scottsdale, AZ, USA) using a Scotch tape (3M Inc. USA) on polydimethylsiloxane (PDMS) gel film (Gel-Pak Inc. Hayward, CA, USA) and the interested few layer flakes were identified under optical microscope, followed by the layer confirmation via Raman characteristics and AFM height profile. Note that, mechanically exfoliated flakes were chosen for their high quality and clean interface promising greater mobility and high performance device fabrication. Further, for the Au contact with $WS_2$, electrodes were patterned in advance on a pre-patterned $SiO_2$/Si (285 nm oxide thickness) substrate using e- beam lithography technique and then Cr (5 nm)/Au (30 nm) were deposited via e-beam evaporation followed by lift-off with acetone. Then, the selected few layer flakes were deterministically transferred on the targeted Au electrodes from PDMS gel film following the dry transfer technique to make sure residue free clean transfer. After the successful transfer of the flakes on Au electrodes, again electrodes were patterned using second step e-beam lithography followed by metal deposition Cr (5 nm)/Au (30 nm) for Cr contacts on $WS_2$. To remove the resist residue and improve the contact conductance, the fabricated devices were annealed at 150°C for 2 hrs in a high vacuum of ~ $10^{-3}$ Torr. Finally, the synthesized diluted solution (0.1 mg $mL^{-1}$) of perovskite NCs was uniformly spin-coated several times (varying from one to four) onto $WS_2$ layer with a speed of 2000 rpm.

**Characterisations and measurements:**

X-ray diffraction (XRD, Philips MRD X-ray diffractometer) patterns were recorded using characteristic Cu-K$\alpha$ ($\lambda$ = 1.5418 Å) radiation with 2.0° grazing incidence angle. For the Transmission electron microscopy (TEM) sample preparation, $CsPbI_3$ NCs solution was dissolved in Hexane and then placed into a TEM grid and dried it for few minutes and did the measurement. The TEM images were carried out using TECNAI G2 TF20-ST and JEM-2100F Field Emission Electron Microscope operating at 200 kV equipped with Gaytan's latest CMOS camera. All the images were proceeding by Digital Micrograph Software for the estimation of d-Spacing & Indexing. UV–vis–NIR absorption spectra of as synthesised $CsPbI_3$ samples were recorded using a fiber probe-based UV–vis–NIR spectrophotometer (Model: U-2910 Spectrophotometer, HITACHI) and a broadband light source. Raman and PL spectra were recorded using a semiconductor laser of excitation wavelength 532 nm, equipped with a CCD detector, an optical microscope of 100x objective lens and a spectrometer (WITec alpha-300R). The photogenerated carrier lifetime was measured by exciting the material with a pulsed diode laser of wavelength 372 nm and detecting the signal using Edinburgh LifeSpec-II fluorescence lifetime spectrometer fitted with a PMT detector. Room-temperature current−voltage characteristics were recorded using a Keithly semiconductor parameter analyzer (4200 SCS) in the presence of an Argon laser (514 nm) and a broadband solar simulator (AM 1.5, 100 mW/$cm^2$) as a visible light source.

## Results and discussion

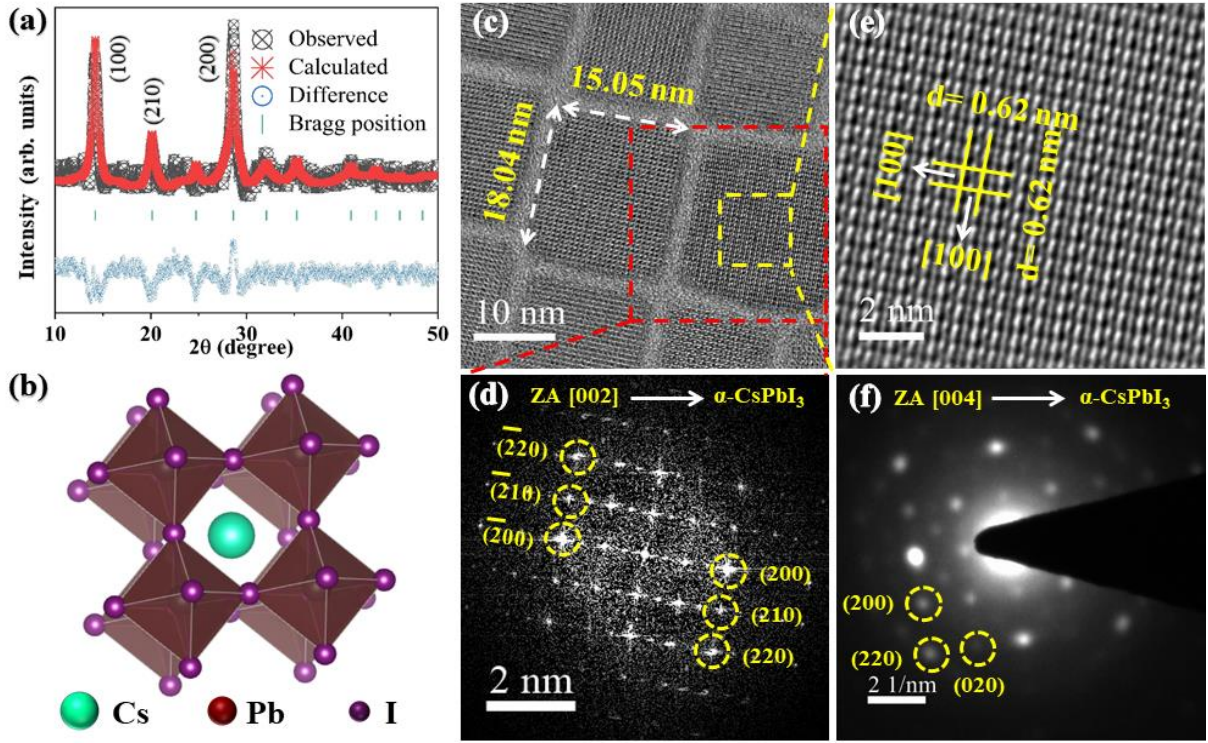

FIG. 1. (a) Rietveld refinements (α-phase fitting) of the XRD pattern of a film of cubic $CsPbI_3$ NCs. (b) 1×1 3D VESTA visualization image of α-$CsPbI_3$ cubic crystal structure. (c) Typical HRTEM image of $CsPbI_3$ NCs with an energy 200 keV revealing cubic morphology. (d) FFT patterns from a region marked by the red dotted square on the micrograph (c). (e) A magnified view of the corresponding HRTEM image in the selected yellow square region on micrograph (c). (f) SAED pattern of α-$CsPbI_3$ NCs showing well defined diffraction spots indexed as (200), (220) and (020) planes viewed along [004] zone axis.

To study the crystal structure of as-synthesized $CsPbI_3$ NCs, we have recorded X-ray diffraction pattern, followed by their fitting with Rietveld refinement full proof software, which are presented in **Fig. 1(a)**. An excellent agreement with fitted results indicates the growth of single-phase (α-phase) cubic $CsPbI_3$. The crystal structure of $CsPbI_3$ NCs visualized using VESTA 3D software through Rietveld fitting is shown in **Fig. 1(b)**. The VESTA 3D (1x1) structure shows the absence of any octahedral inclination in the perovskite NCs, which is known to be beneficial for achieving higher stability under laboratory ambient (45-50% humidity). Typical high resolution transmission electron microscopy (HRTEM) image reveals almost cubic shape of the synthesised NCs (15.05 nm × 18.04 nm), as shown in **Fig. 1(c)**. Corresponding first Fourier transform (FFT) pattern presented in **Fig. 1(d)** of the red squared portion of the **Fig. 1(c)**, shows pure cubic α-phase pattern along the zone axis [002]. Whereas, the high-resolution fringe pattern from the yellow squared region of **Fig. 1(c)** shows a d-spacing of 0.62 nm, which is in well matched with the cubic α-phase of $CsPbI_3$ [30], as shown in **Fig. 1(e)**. The result indicates (100) directional growth of cubic phase $CsPbI_3$ NCs, which is in well agreement with our previously reported results [30]. Corresponding selected area electron diffraction (SAED) patterns shown in the **Fig. 1(f)**, with indexed (200), (220) and (020) planes along the zone axis [004], also corroborate the pure cubic structure of synthesized $CsPbI_3$.

**Figure 2(a)** presents the optical absorption and emission properties of the as-synthesised α-phase $CsPbI_3$ NCs in the visible wavelength range with an absorption maxima at ~ 680 nm and the corresponding bandgap value is ~ 1.814 eV [30], extracted from the Tauc plot shown in Fig. S1 within the Supplimental Material. Further, the photoluminescence (PL) maxima at ~ 687 nm confirms the formation of excitons ($X_α$) across the direct bandgap (~1.80 eV) of α-phase $CsPbI_3$ NCs represented via blue curve in **Fig. 2(a)**. The Gaussian line shape of the PL spectrum and the absence of any other PL peaks clearly dictate the synthesis of pure α-phase $CsPbI_3$ without presence of any mixed phase. To examine the charge transfer mechanism at the $CsPbI_3$ NCs/$WS_2$ interface, Raman spectroscopy and micro-PL (μ-PL) measurements have been carried out by spin-coating of a dilute solution of $CsPbI_3$ NCs uniformly on the exfoliated $WS_2$ surface. For the room temperature μ-Raman-PL measurements, samples have been excited with a CW laser having a wavelength of 532 nm with the laser power being kept at a very low value to avoid any local heating induced sample degradation. **Figure 2(b)** represents comparative Raman spectra of $WS_2$ and mixed dimensional van der Waals heterostructure (MvWH) samples, showing intense in-plane $2LA+E^1_{2g}$ Raman modes at ~ 351 cm$^{-1}$ and out-of-plane vibrational $A_{1g}$ peaks at ~ 420 cm$^{-1}$ [32]. The $A_{1g}$ vibrational Raman mode, which preserves the symmetry of the lattice, is clearly red-shifted by ~ 4.7 cm$^{-1}$ in case of $CsPbI_3$ decorated $WS_2$ layer [**inset of Fig. 2(b)**], revealing the interfacial charge transfer phenomena.

The external electron doping in 2D WS$_2$ leads to the filling-up of antibonding states of the conduction band, mostly made up of d$_{z^2}$ orbitals of transition metal atoms [33]. This makes the bonds weaker and the A$_{1g}$ peak of pristine WS$_2$ is shifted towards a lower wavenumber on significant electron doping from CsPbI$_3$ NCs [34].

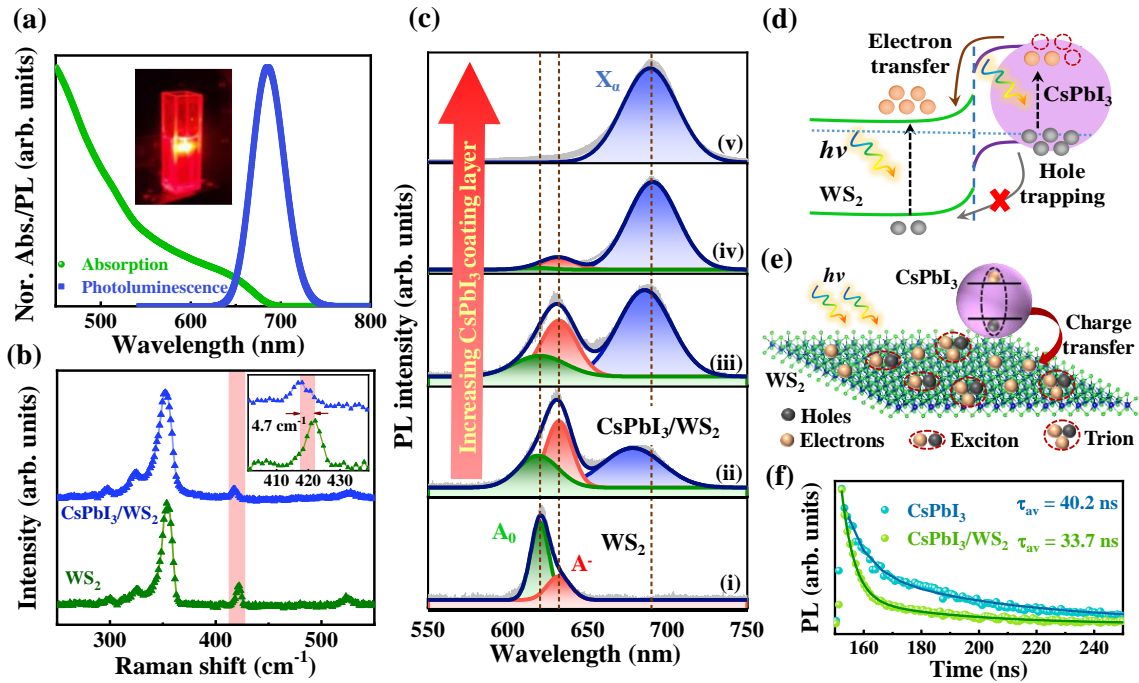

FIG. 2. (a) Absorption (Green) and photoluminescence (Blue) spectra of as-synthesised cubic phase CsPbI$_3$ NCs. (b) Comparative Raman spectra of WS$_2$ flakes before and after CsPbI$_3$ NCs decoration showing characteristic E$_{2g}$ and A$_{1g}$ peaks of layered WS$_2$. Inset shows the magnified image of the out-of-plane A$_{1g}$ mode revealing a clear peak shift to lower wavenumbers due to electron doping in WS$_2$ flakes from CsPbI$_3$ NCs. (c) Deconvoluted PL spectra of (i) a bare ML WS$_2$ flake and (ii-v) the heterostructure samples with varying number of spin coated layers of CsPbI$_3$ NCs on the WS$_2$ flake. The spectra (ii), (iii), (iv) and (v) represent the PL emission from the first, second, third and fourth spin coated layers of CsPbI$_3$ NCs, respectively. The green (red) peak represents the A excitonic (A$^-$ trionic) emission from ML WS$_2$ flakes and the blue peak represents the emission from band to band transition of cubic α-phase CsPbI$_3$ NCs. (d) Energy band diagram of CsPbI$_3$/WS$_2$ hybrid heterostructures showing effective electron doping in WS$_2$ from CsPbI$_3$ sensitizers and hole trapping in the NCs giving rise to a strong photogating effect. (e) Schematic representation of the charge transfer mechanism in 0D/2D CsPbI$_3$/WS$_2$ hybrid heterostructures giving rise to trion formation in ML WS$_2$. (f) Normalised time resolved PL decay curves of CsPbI$_3$ NCs (Blue curve) and CsPbI$_3$/WS$_2$ hybrids (Red curve), measured using an excitation wavelength of 372 nm.

On the other hand, the monolayer (ML) WS$_2$ PL emission characteristics [**Fig. 2c(i)**] consist of a strong A-excitonic emission at ∼ 1.995 eV, corresponding to the direct band-to-band transitions at the K (and/or K') point of the Brillouin zone, and a weaker trionic A$^-$ emission at ~ 1.965 eV with a binding energy of ~ 30 meV, which are in good agreement with the previously reported results [34]. It is to be noted that, ML WS$_2$ is purposefully chosen for the charge transfer study via PL measurements due to its extraordinary luminescence property at room temperature owing to the direct bandgap transition. The existence of trions in the room temperature emission spectrum indicates the unintentional doping in the un-passivated WS$_2$ flake from the substrate as well as the surrounding environment [35,36]. A systematic PL study of the hybrid structure with increasing layer numbers of CsPbI$_3$ NCs spin-coated over ML WS$_2$ shows a pronounced excitonic-PL quenching in MvWH as compared to both the pristine materials (Fig. S2 within the Supplemental Material). Further, the CsPbI$_3$/WS$_2$ MvWHs show relatively broad PL spectra with combined contributions from CsPbI$_3$ NCs as well as ML WS$_2$ and the spectral shape changes with increasing CsPbI$_3$ spin-coated layer numbers **[Figs. 2c(ii-v)].** To explore the effect of CsPbI$_3$ coating over WS$_2$, we have fitted each spectrum with three Gaussian peaks containing the characteristics excitonic features of both the materials and analysed their intensity variation with increasing concentration of CsPbI$_3$ treated on WS$_2$ flakes, as shown in **Fig. 2(c)**. It is noticed that the distinctive trion peak (A$^-$) of WS$_2$ becomes prominent with increasing density (coating number) of CsPbI$_3$ NCs and finally excitonic to trionic (integrated intensity) crossover is observed above a critical concentration of CsPbI$_3$ NCs (Fig. S3 within the Supplemental Material). The possible explanation behind these observations is as follows: upon illumination, photoexcited electron-hole pairs are generated in both WS$_2$ and CsPbI$_3$, however, due to the type-II energy band alignment of the heterostructures, electrons are easily transferred from CsPbI$_3$ NCs to ML WS$_2$, as illustrated in **Fig. 2(d)**. On the other hand, photogenerated holes remain trapped in the NCs, resulting in a reduced recombination rate and giving rise to a quenched excitonic PL intensity for CsPbI$_3$ NCs and increased trionic emission in ML WS$_2$. The enhanced generation rate of trions results in the reduced density of excitons inside the system, leading to the suppression of excitonic peak intensity and the dominance of the trion peak in the PL spectra of MvWH, as schematically depicted in **Fig. 2(e)**. [34] To further confirm the charge transfer phenomena, time-resolved photoluminescence (Tr-PL) spectra have been measured. **Fig. 2(f)** shows the Tr-PL decay curves of $CsPbI_3$ NCs (blue curve) and $CsPbI_3/WS_2$ MvWHs (red curve). The PL decay curves have been fitted using a bi-exponential function to extract the average excitonic life time ($\tau_{av}$). The $\tau_{av}$ of $CsPbI_3$ NCs decreases from 40.2 ns to 33.7 ns after hybridization with $WS_2$, corroborating the successful charge transfer mechanism from $CsPbI_3$ NCs to $WS_2$ flakes. As a conclusion, the type-II band alignment in $CsPbI_3/WS_2$ heterostructure facilitates an efficient electron–hole pair separation and strong electron doping into $WS_2$ channel, making the hybrid system ideal for fabrication of superior performance phototransistor devices [37].

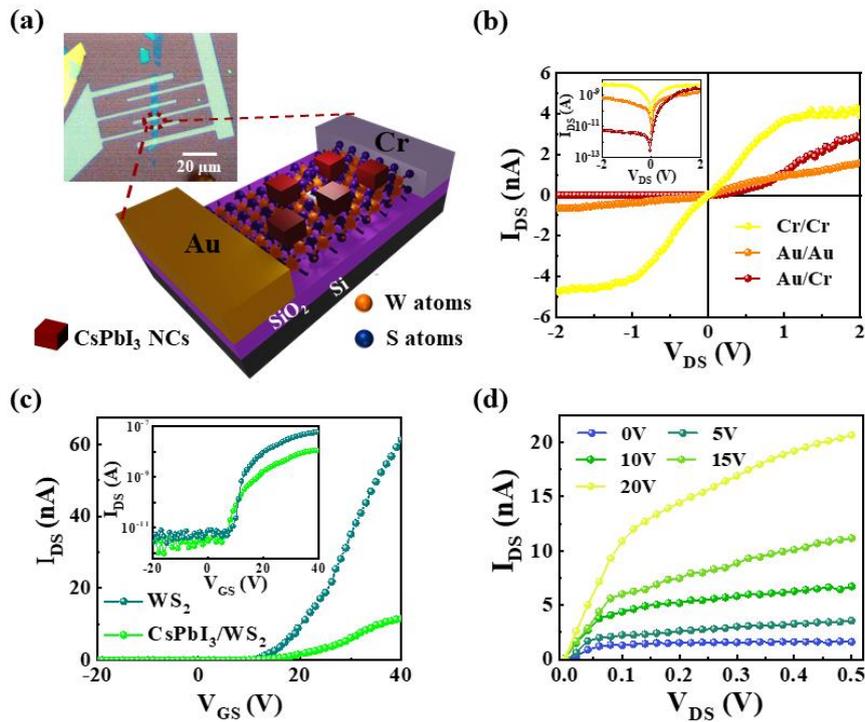

FIG. 3. (a) Schematic 3D view of the fabricated back gated phototransistor comprising of $CsPbI_3$ sensitized $WS_2$ channel with asymmetric electrodes (Au and Cr) acting as a source and drain. An optical micrograph of the device is shown in the inset. (b) Linear $I_{DS}$-$V_{DS}$ characteristics of three fabricated devices with different source-drain contacts (i) Cr-Cr (Yellow curve), (ii) Au-Au (Orange curve) and (iii) Au-Cr (Brown curve) without any back gate bias. Inset: the corresponding semi-logarithmic $I_{DS}$-$V_{DS}$ characteristics plots. (c) Transfer

($I_{DS}$-$V_{GS}$) characteristics of the CsPbI$_3$/WS$_2$ hybrid transistor (Green curve) and control WS$_2$ transistor (Blue curve) devices in linear scale and logarithmic scale (Inset). The current value in the accumulation region decreases and the threshold voltage is shifted to a higher positive voltage in hybrid device due to charge transfer through the CsPbI$_3$/WS$_2$ junction. (d) Output characteristics of the hybrid device with varying gate voltage.

The CsPbI$_3$ NCs/WS$_2$ MvWH photo-FET with asymmetric metal contacts has been demonstrated by exploiting the Schottky barrier induced dark current suppression for the zero gate bias driven photosensitivity of the device with a simpler fabrication technique. **Figure 3(a)** schematically demonstrates the as-fabricated phototransistor device structure consisting of a few layer WS$_2$ channel and asymmetric Cr and Au electrodes as source and drain terminals, respectively. The inset shows the optical micrograph of the connected few layer WS$_2$ flake (5×2 μm$^2$) having a thickness of ~ 4.5 nm corresponding to 4-5 atomic layers of WS$_2$, further corroborated by the AFM analysis (Fig. S4 within the Supplemental Material). The deposition of a lower work function Cr ($\Phi_{Cr}$ = 4.5 eV) and higher work function Au ($\Phi_{Au}$ = 5.1 eV) on n-type WS$_2$ as asymmetric contacts reveals room temperature rectifying diode characteristics with rectification ratio up to 5.2 × 10$^2$ even at zero applied back gate bias, as depicted via the current-voltage ($I_{DS}$-$V_{DS}$) characteristics in **Fig. 3(b)**. Under an applied reverse drain voltage, the potential barrier height between Au and WS$_2$ becomes higher to suppress the current flow through the junction compared to Cr, and thus exhibits the $I_{DS}$-$V_{DS}$ characteristics of an ideal diode [13]. The current through the Au–WS$_2$-Cr device follows the diode behaviour and the Schottky barrier ($\Phi_b$) at Au-WS$_2$ interface is capable of reducing the dark current significantly making the device architecture an ideal prototype for operating in full depletion mode even at zero gate bias, leading to a very high ON-OFF ratio of the device. On the other hand, a linear $I_{DS}$-$V_{DS}$ characteristics with a comparatively larger current value (100 nA) confirms the formation of an Ohmic-like junction with a very low contact resistance for Cr-WS$_2$-Cr device [38]. Further, Au-WS$_2$-Au system reveals a rectifying output characteristics with relatively lower current than Cr contacts, confirming a typical high resistive back-to-back

Schottky diode [15]. The energy band alignment with different metal contacts is schematically depicted in Fig. S5 within the Supplemental Material, revealing an easy current flow through the Ohmic Cr junction and restricted flow via built-in potential barrier in the Au Schottky junction. **Fig. 3(c)** shows the transfer ($I_{DS}$-$V_{GS}$) characteristics of the $WS_2$ phototransistor with asymmetric Cr–Au contacts at a reverse drain voltage of -2V revealing excellent n-type channel properties at room temperature with off-currents of the order of 10 pA and the transistor ON-OFF ratio ~$10^4$. Further, to understand the effect of $CsPbI_3$ treatment on the device performance, the $CsPbI_3$/$WS_2$ hybrid transistor characteristics is compared with the pristine $WS_2$ one, referred to as the control device. The incorporation of sensitizing perovskite NCs on 2D-$WS_2$ layer results in a junction formation via Fermi level alignment in equilibrium under dark condition. In this process, the draining of electrons from the $WS_2$ channel towards $CsPbI_3$ NCs results in the depletion of majority carriers in $WS_2$ leading to the lowering of the current flow in the channel under dark condition. Further, we have studied the output characteristics of the hybrid phototransistor on application of gate voltage varying from 0 to +20 V, as depicted in **Fig. 3(d)**. For a higher positive gate voltage, more electrons are induced in the $WS_2$ channel and the transistor has a higher current in the saturation state. On the other hand, under a negative gate bias a small amount of current flows through the channel owing to the depletion of carriers, leading to the OFF state of the transistor [**Fig. 3(c)**].

The performance of the $CsPbI_3$/$WS_2$ MvWH photo-FET has been analysed by recording the room temperature $I_{DS}$-$V_{DS}$ characteristics for zero gate bias under dark as well as visible illumination using a Newport solar simulator having broadband emission with irradiance of 100 mW/cm$^2$ under air mass (AM) 1.5G condition, as shown in **Fig. 4(a)**. For comparison, the characteristics of the pristine $WS_2$ control device is also presented. The suppression of dark current to the order of tens of pA even without any gate bias along with the significant reduction of noise currents are attributed to the built-in electric field at the Au/$WS_2$ Schottky junction, which further helps in effective separation of photogenerated carriers created in $WS_2$ channel. On illuminating the heterojunction device, the reverse current tends to increase due to the collection of photogenerated minority carriers (holes) at the electrodes. The photo-to-dark current ratio of MvWH photo-FETs by illuminating with a broadband light source is estimated to be much higher compared to the pristine $WS_2$ one (~1000 times) under the applied reverse bias condition, which reaches to a value of ~$1.08 \times 10^6$ at $V_{DS}$ of -2V, as shown in Fig. S6 within the Supplemental Material. The decoration of $WS_2$ channel with superior light absorbing perovskite $CsPbI_3$ NCs facilitates enhancement in the photocurrent by elevating the photogenerated carriers in the channel via efficient charge transfer from $CsPbI_3$ to $WS_2$ due to type-II energy band alignment [7,19]. This explains the significant enhancement (~$10^3$ times) of response of MvWH transistor over the control one, revealing the role of photoabsorbing $CsPbI_3$ NCs in boosting the performance of the phototransistor. Further, the spectral responsivity of the fabricated MvWH photo-FET, the most important figure of merit to evaluate a detector performance, has been studied displaying a broadband spectral photoresponse covering the entire visible wavelength range, as shown in **Fig. 4(b)**. It may be noted that a peak responsivity of ~$1.05\times10^2$ A/W at ~ 460 nm at an applied bias ($V_{DS}$) of -2 V is achieved, which is close to the C-exciton absorption edge of $WS_2$. Two other peaks at ~ 620 nm (R ~ $0.97\times10^2$ A/W) and ~ 720 nm (R ~ $0.77\times10^2$ A/W) correspond to the direct bandgap absorption of few layer $WS_2$ and α-phase $CsPbI_3$ NCs, respectively. Further, the spectral responsivity increases with increasing reverse $V_{DS}$ that assists in the efficient extraction of photogenerated carriers. On the other hand, the control $WS_2$ based device also exhibits a similar trend with increasing bias showing a maximum peak responsivity of ~ 10 A/W at ~ 460 nm at -2 V applied $V_{DS}$ (Fig. S7 within the Supplemental Material). It is to be noted that the decoration of $CsPbI_3$ NCs on $WS_2$ flakes not only improves the detector responsivity by more than 10-fold but also extends the spectral responsivity window up to 800 nm, as shown comparatively in **Fig. 4(b)**. Hence, the decoration of $WS_2$ active channel layer with excellent photoabsorbing $CsPbI_3$ NCs appear to be a promising approach for next generation high performance optoelectronic applications. To further investigate the role of $CsPbI_3$ in photocarrier generation and efficient charge transfer, the electromagnetic simulations have been performed using the COMSOL Multiphysics software. **Figure 4(c)** shows the electric field distribution of the hybrid $CsPbI_3/WS_2$ device, illuminated with an electromagnetic plane wave of wavelength λ = 680 nm from top, which propagates through air and the nanostructure. The distribution clearly depicts that the electric field is trapped along the edges of the nano-cubes of $CsPbI_3$ with the maximum confinement occurring near the base (as demonstrated by the colour index profile), resulting in strong charge transport in $CsPbI_3/WS_2$ hybrid heterostructure.

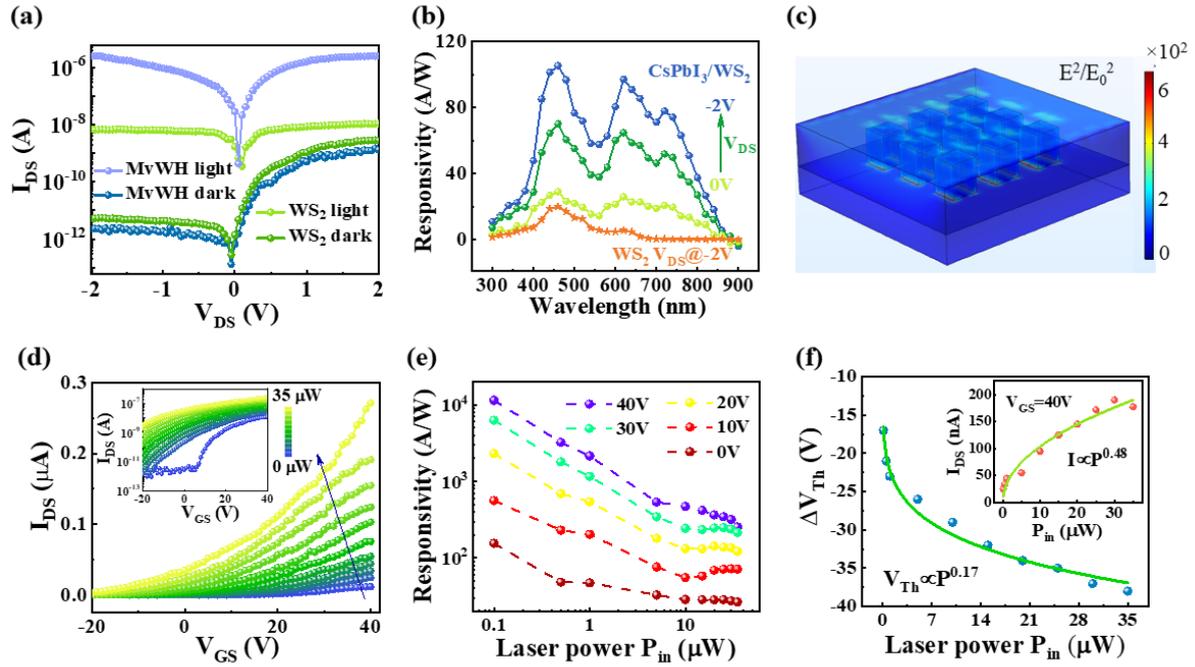

FIG. 4. (a) Comparative $I_{DS}$-$V_{DS}$ characteristics of pristine $WS_2$ and $CsPbI_3/WS_2$ MvWH photo-FET under dark and illumination via broadband light source for zero gate bias. (b) Spectral responsivity curves of MvWH photo-FET at $V_{GS} = 0V$ with increasing reverse $V_{DS}$ from 0V to -2V, as shown via yellow, green and blue curves. The blue curve represents the responsivity of the photo-FET at a maximum $V_{DS}$ of -2V, while the spectral responsivity of the control device (orange curve) showing an order of magnitude lower device response at same $V_{DS}$. (c) COMSOL Multiphysics simulated E-field distribution at the vicinity of the hybrid system upon excitation with an excitation wavelength of 680 nm. (d) The transfer characteristics ($I_{DS}$-$V_{GS}$) of MvWH photo-FET for a range of incident powers (from 0.1 to 35 μW) with an illumination of wavelength 514 nm at $V_{DS} = -2$ V. (e) The variation of responsivity of the device with incident illumination power for $V_{GS}$ varying from 0 to 40 V. (f) The shift in threshold voltage ($\Delta V_{Th}$) with increasing illumination power ($P_{in}$) fitted with a power law. Blue dots represent the extracted data points from panel (a) and green line represents the fitted curve. Inset: The power law fit of the variation of photocurrent with incident power at $V_{DS} = -2V$ and $V_{GS} = 40V$ showing a sublinear photocurrent dependency with incident optical power.

Further to explore the impact of gate bias on transistor performance, the photo-induced transfer characteristics ($I_{DS}-V_{GS}$) of the $WS_2/CsPbI_3$ MvWH photo-FET is recorded under the dark (black markers) and 514 nm illumination with a range of optical powers (from 0.1 μW to 35 μW) at a constant $V_{DS}$ of -2V, as illustrated in **Fig. 4(d).** The corresponding logarithmic current representation is depicted in the inset. Under illumination of a fixed power of 35 μW, the drain current of the MvWH photo-FET is enhanced by ~ 13 times (from ~ 20 nA to ~ 0.26 μA) at a constant gate voltage of ~ 40V, manifested by the strong photoabsorption in $CsPbI_3$ and subsequent transfer of photoexcited electrons to the $WS_2$ channel. Further, with the increase of laser power, the photocurrent significantly increases in the accumulation region (i.e. $V_{GS}>V_{Th}$) and the transfer curves are gradually shifted to a negative gate voltage. As illustrated in **Fig. 2(d),** the favourable energy band alignment rules out the possibility of hole injection from $CsPbI_3$ into $WS_2$, leading to the trapped holes induced strong photogating effect in the hybrid system. This leads to significant photocurrent increment in the accumulation region and negative threshold voltage shift ($\Delta V_{Th}$) with increasing incident power density of illumination [39]. To investigate in greater detail, the calculated responsivity as a function of illumination power has been plotted for different gate bias voltages in **Fig. 4(e)**. Here, the responsivity value increases with increasing positive gate bias in case of MvWH photo-FETs and reaches to a high value of ~ $1.1 \times 10^4$ A W$^{-1}$ at a back gate voltage of ~ 40 V under an illumination power of 0.1 μW, which is quite remarkable compared to those previously reported 0D/2D hybrid phototransistors [37,39]. Note that, for all the gate voltages, the measured responsivity dropped with increasing power because of the saturation of sensitizing traps in $CsPbI_3$ NCs, which is a characteristic footprint of trap-dominated photoresponse [40–

42]. Further, we have extracted the threshold voltage via extrapolating the linear region of each transfer curve under different incident laser powers and the shift in threshold voltage is plotted as a function of incident power. The variation is fitted with the power law function $V_{Th} \propto P^b$, as depicted in **Fig. 4(f),** to understand the possible photoconduction mechanism. The extracted fitting exponent, b ~ 0.17 clearly indicates a sublinear dependency on laser power confirming the existence of photogating dominant carrier conduction in MvWH photo-FETs [43]. Further, it is also observed that the change in $V_{Th}$ is large in the lower power region and starts to saturate gradually at a higher power owing to the saturated trap states present in sensitizer interface leading to the saturation of the photogating effect. The photocurrent $I_{Ph} = I_{Photo} - I_{Dark}$ versus gate voltage for different illumination intensity [**Fig. 5(a)**] shows a strong modulation with $V_{GS}$, and a clear maximum in response can be identified around +35 V. The strongest response of the FET device corresponds to the region with highest transconductance, due to the favourable Fermi level alignment, for low-contact resistance operation leading to many cycles of electron circulation to produce maximum gain. Hitherto, in this region the FET device operates at a relatively higher dark current, compromising the signal-to-noise ratio (SNR) of the device, which is also a very important figure of merit of photo-FETs. The SNR defined as $I_{Photo}/I_{Dark}$ is illustrated in the same panel, **Fig. 5(a)**, which reveals the potential of 0D/2D hybrid phototransistors for highest sensitivity detection in its depletion regime with $V_{GS}$ from 0 to 5V. In this region, a lowest dark current and a maximum sensitivity are achieved, despite the devices' concurrent drop in the photocurrent. So the maximum sensitivity of the device can be achieved via contact engineering where the transistor is operated in the depletion region, even without applying any gate bias, hitherto unreported for photo-FET devices. While the peak responsivity of our device is comparable or superior to the reported 2D materials based hybrid phototransistor devices with perovskite sensitizers, the sensitivity is found to be significantly higher without application of any external gate bias (see **Table 1**). These results illustrate the superior performance of broadband phototransistor, with ultrahigh sensitivity and responsivity, using CsPbI$_3$ NCs sensitized 2D WS$_2$ layer.

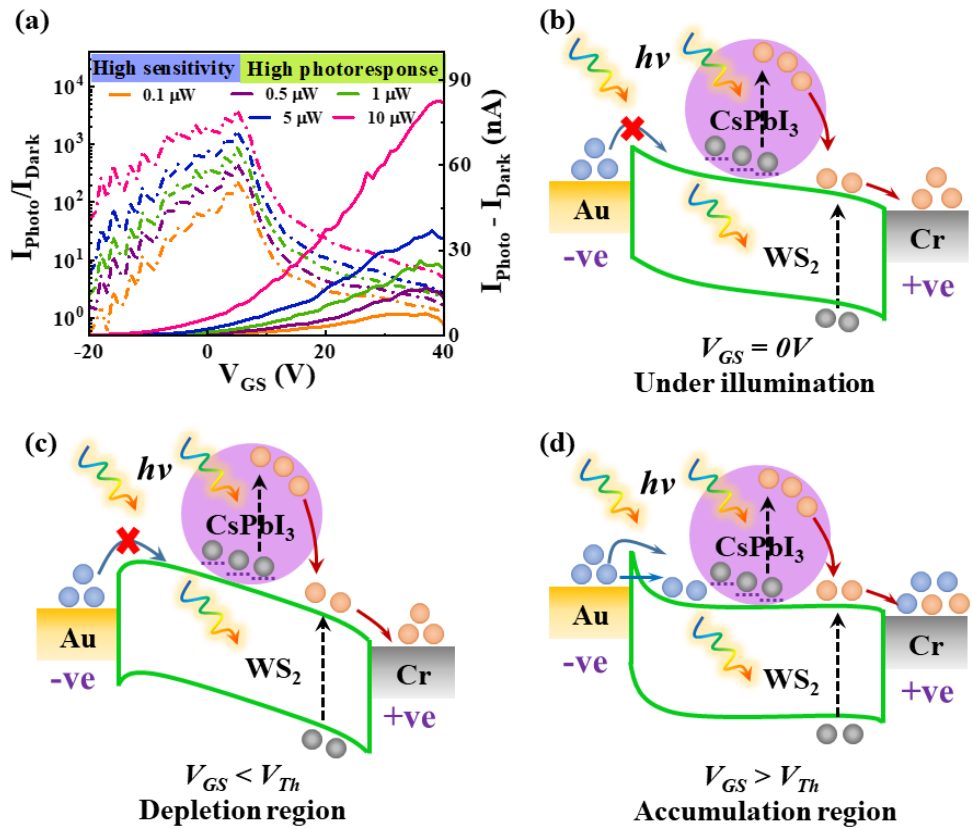

FIG. 5. (a) Back-gate bias dependent photocurrent (right axis) and photo-to-dark current ratio, (left axis) of the phototransistor device under five different illumination intensities (from 0.1 μW to 10 μW) for 514 nm. Despite the strongest photoresponse at higher gate bias (V$_{GS}$ ≈ 40 V), highest sensitivity of the device is achieved in the depletion regime (V$_{GS}$ ≈ 0V). The schematic representation of channel current transport mechanism and energy band diagram of the asymmetric contact hybrid phototransistor under reverse drain-source voltage with (b) zero and (c-d) different gate bias conditions.

On the other hand, a remarkable photoresponse of CsPbI$_3$/WS$_2$ MvWH photo-FET is explained by considering the influence of positive gate voltage on energy band alignment at the contact interfaces and heterostuctures leading to efficient charge injection into n-type WS$_2$ channel layer from photoabsorbing CsPbI$_3$ NCs. As illustrated in **Fig. 5(b)**, the Schottky barrier at the Au/WS$_2$ interface is high enough to inhibit the charge conduction mechanism across the WS$_2$ channel layer at reverse drain bias without any gate electric field under dark condition. Hence, the transistor immediately goes to the OFF state with very low dark current in the order of pA. At this condition, when the visible light is illuminated on the 0D/2D heterostructure, the photogeneration takes place in both CsPbI$_3$ NCs as well as WS$_2$ channel layer, as depicted in **Fig. 5(b)**. The effective photogenerated carrier separation takes place by the built-in electric field at the Schottky junction (WS$_2$/Au) as well as at CsPbI$_3$/WS$_2$ interfaces. The subsequent transition of photoexcited electrons from CsPbI$_3$ to WS$_2$ starts to populate the active channel layer which are collected by the external electrodes under an applied reverse $V_{DS}$, leading to the photoresponsivity of ~ $10^2$ A/W at $V_{DS} = -2V$ and $V_{GS} = 0V$. Further, the application of a back gate voltage ($V_{GS}$) to the device modulates the Schottky barrier height at Au/WS$_2$ interface as shown in **Figs. 5(c)-(d)** [52,53]. An application of negative gate bias ($V_{GS} < V_{Th}$) increases the barrier height leading to the transistor operation in the depletion region, where the photosensitivity ($I_{Photo}/I_{Dark}$) of the device is maximum. On the other hand, on increasing the $V_{GS}$ beyond $V_{Th}$ initiates the lowering of the Schottky barrier at Au/WS$_2$ interface, resulting in a higher magnitude of charge carrier injection from the Au electrode to WS$_2$ channel layer through thermoionic as well as tunnelling mechanisms, as illustrated in **Fig. 5(d)**. Thus, the cumulative effects of CsPbI$_3$ NCs decoration mediated strong photogating phenomena as well as the gate voltage induced Schottky barrier lowering result in a drastic enhancement of the photocurrent ($I_{Photo} - I_{Dark}$) through the transistor channel at ON state ($V_{GS} > V_{Th}$). This leads to an ultrahigh photoresponsivity of the order of ~ $10^4$ A/W at $V_{GS} = 40$ V. Such gate modulated responsivity and sensitivity of MvWH photo-FET devices via interface engineering offers a novel pathway for next generation high performance and low power integrated photonic technology.

**Table 1.** Comparison of device performances with reported 2D material based hybrid photo-FETs with perovskite sensitizers

| Device structure | Sensitizer | Operational spectral range | $I_{dark}$ w/o applied $V_{GS}$ | $I_{photo}/I_{dark}$ @ $V_{GS}=0V$ | Responsivity for different values of $V_{Gs}$ | Ref. |
|---|---|---|---|---|---|---|
| Au / ML WS$_2$/ Au | CH$_3$NH$_3$PbI$_3$ | 450-700 nm | 5nA | $10^4$ | 2.5 A/W @ 0V | [44] |
| Au / ML MoS$_2$/ Au | CsPbBr$_3$ | 350-550nm | 0.2 nA | $10^3$ | 4.4 A/W @ 0V | [45] |
| Au / Ti / ML MoS$_2$ / Ti / Au | Ch$_3$NH$_3$PbBr$_3$ /CsPbI$_{3-x}$Br$_x$ | 532 and 355 nm | 4 nA | $10^4$ | $7 \times 10^4$ A/W @ 60V | [46] |
| Au / Ti / FL MoS$_2$/ Ti / Au | CsPbBr$_3$ | 405 nm | 10 nA | 10 | $4.7 \times 10^4$ A/W @ 20V | [47] |
| Au / FL BP / Au | CsPbBr$_3$ | 405 nm | 2 nA | $10^2$ | 357.2 mA/W @ 0V | [48] |
| Au / ML MoS$_2$/ Au | CsPbI$_{3-x}$Br$_x$ | 532 nm | 0.2 μA | $10^3$ | $1.13 \times 10^5$ A/W @ 60V | [49] |
| Au / FL BP / Al / Au | MAPbI$_{3-x}$Cl$_x$ | 400-900 nm | 0.1 μA /μm | $10^2$ | $4 \times 10^6$ A/W @ 40V | [50] |
| Au / Ti / FL MoSe$_2$(WSe$_2$) / Ti / Au | CsPb(Cl/Br)$_3$ | 455 nm | 0.8 nA | 10 | $10^2$ A/W @ 50V | [51] |
| Au / Cr / FL Ta$_2$NiSe$_5$/ Cr / Au | CH$_3$NH$_3$PbI$_3$ | 800 nm | 3.5 μA | 10 | $2.4 \times 10^2$ A/W @ 0V | [37] |
| **Au / Cr / FL WS$_2$/ Au / Cr** | **α-phase CsPbI$_3$** | **400-800 nm** | **2 pA** | **$10^6$** | **$10^4$ A/W @ 40V** | **Our work** |

Temporal photoresponse is also an important parameter for the phototransistors performance in terms of switching speed and device stability. The transient photoresponse of the as-fabricated CsPbI$_3$/WS$_2$ MvWH photo-FET upon visible illumination ($\lambda = 514$ nm) at $V_{GS} = 0V$ with varying reverse $V_{DS}$ is demonstrated in **Fig. 6(a)**. Upon illumination of four periodic pulses of the Argon-ion laser, relatively fast and consistent photocurrent modulation characteristics of the device reveals the stability and reproducibility of the as-fabricated MvWH photo-FET. The device exhibits a much stronger photoresponse characteristics revealing ratio of ~$10^6$ as compared to the control device with pristine WS$_2$ with the value ~$10^3$ (Fig. S8 within the Supplemental Material), which is attributed to the injection of high density photogenerated charge carriers into the $WS_2$ channel from strong light absorbing $CsPbI_3$ NCs. With the increment of reverse $V_{DS}$, a consistent photocurrent enhancement is distinctly noticed from the switching characteristics owing to the increase of depletion region width at the Schottky barrier interface and subsequent separation of photogenerated charge carriers. Further, the rise and fall times of the fabricated device in the absence of gate bias have been estimated using an enlarged single cycle response [**Fig. 6(b)**] and are found to be around ~ 43.8 ms and ~ 32.7 ms, respectively, which are further reduced to 34 ms and 24 ms, respectively on applying a gate voltage of 40 V. The response speed of these devices are found to be relatively slower, which is attributed to the trapping of charge carriers in various structural and surface defect states present in $WS_2$ as well as $CsPbI_3$ NCs and their local junction interfaces. These interface traps present in $WS_2$ layer are mostly empty when biased under depletion condition, i.e. $V_{GS} < V_{Th}$ owing to lack of enough mobile carriers in the channel. This allows a large number of photogenerated electrons to get trapped by the defect states while some of the gate induced electrons, although small in number, can be trapped as well. This results in a relatively slow rise of current, as depicted in the photocurrent dynamic response. On the other hand, the interface traps are nearly filled up with gate-induced electrons in accumulation condition, when $V_{GS} > V_{Th}$, as shown in **Fig. 6(b)**. Hence, the trapping probability of photogenerated carriers is lower and a relatively faster response (~34 ms) is observed in MvWH photo-FET devices. Further, owing to the fact that the perovskite materials are prone to environmental degradation via oxygen diffusion through iodide vacancies upon illumination, the long term operation stability of the fabricated devices have been tested in this study upon visible light illumination at zero gate bias for prolonged duration (more than 60 min). From the I–t curves for the first 100 s [**Fig. 6(c),** left] and the last 100 s [**Fig. 6(c)**, right], it is observed that the photocurrent has almost no attenuation, indicating that these devices show an excellent light stability under ambient condition, even without the use of a glovebox or encapsulation. The device stability has also been tested via recording the photocurrent under illumination over a period of one week, as illustrated in **Fig. 6(d)**. Here, the phototransistor sustains under laboratory ambient conditions (relative humidity (RH) ~ 45-50%, temperature ~ 22°C) for one week with negligible change in the photocurrent via degradation after storing. Further, as $CsPbI_3$ NCs are vulnerable to environmental humidity, to explore the device performance in the extreme humid condition, we have performed the temporal response under 65% RH showing an insignificant degradation (5% decay) in terms of device response [**Fig. 6(e)**]. This superior performance stability is due to the surface defect passivation of $CsPbI_3$ through the interaction with the sulphur of $WS_2$ ensuring the outstanding environmental stability of as-fabricated CsPbI$_3$/WS$_2$ MvWH photo-FETs. The sulfur atoms present on the top layer of WS$_2$ may have stronger coordination to the Pb$^{2+}$ centers of CsPbI$_3$ NCs leads to reduced defect states in perovskites enabling higher reluctance to the degradation [31]. It may be noted that the performance of the devices could be further improved by process optimization, device encapsulation and incorporation of buffer layers. This work reveals the significant potential of colloidal synthesized air-stable α-CsPbI$_3$ NCs on 2D materials in fabricating 0D/2D mixed-dimensional heterostructure photo-FETs for applications in next generation optoelectronic devices.

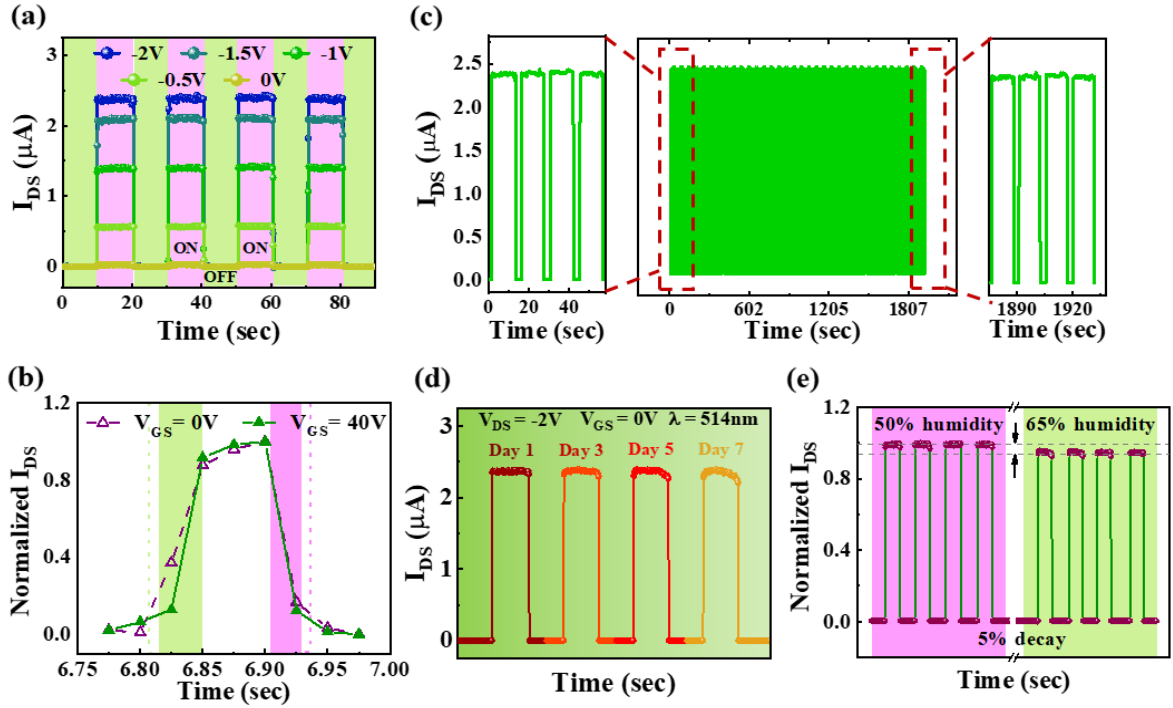

FIG. 6. (a) Transient response of the MvWH photo-FET device under illumination of a 514 nm laser at different applied $V_{DS}$. (b) Temporal photocurrent response of the MvWH device for a wavelength of 514 nm with and without any applied gate bias. The temporal response indicates a significant decrease in rise time (from 43.8 to 34 ms) as well as fall time (from 32.7 to 24 ms), measured at a relatively higher power of 35 µW. (c) Operational stability of the fabricated MvWH photo-FET device under visible illumination for more than half an hour. (d) The transient photocurrent response of the fabricated transistor over a span of seven days from the beginning and end of the stability test. (e) Stability of the device tested under extreme humid conditions (varying from 50 to 65% RH). The last four cycle is the response under 65% of humidity showing around 5% decay in the photoresponse.

**Conclusion:**

To summarize, significant improvements in performance have been realized in CsPbI$_3$/WS$_2$ 0D/2D mixed-dimensional phototransistors with asymmetric metal electrodes leading to combinatorial effect of Schottky barrier induced suppression of dark current and efficient charge transfer from photoabsorbing CsPbI$_3$ nanocrystals, resulting in enhanced photosensitivity and spectral responsivity. The WS$_2$ channel with asymmetric contacts (Cr/WS$_2$/Au) shows a rectifying I-V characteristics under an applied V$_{DS}$ with the dark current in the order of pA. Further, by combining the channel sensitization via decorating the WS$_2$ with photosensitive air-stable α-phase CsPbI$_3$ NCs, a responsivity of ~10$^2$ A/W has been achieved at low V$_{DS}$ (~ -2V) for an incident optical power of 0.1 µW even without any external gate bias. The device exhibits a broad spectral photoresponsivity between 400 and 800 nm due to the extended visible light absorption features of CsPbI$_3$ NCs. Using gate-controlled carrier modulation in the transistor channel, a peak responsivity ~10$^4$ A/W (V$_{GS}$ = +40 V) has been achieved owing to the photogating effect mediated charge conduction whereas the maximum sensitivity (~ 10$^6$ at ~ V$_{DS}$ = -2 V) in terms of signal-to-noise ratio is observed by depleting the channel carries (V$_{GS}$ = 0 to 5 V). These devices show superior performance in terms of environment stability, owing to the filling of surface trap states present in CsPbI$_3$ NCs via conjugation with sulfur atoms of 2D WS$_2$ layer. The fabricated hybrid heterostructure devices combining 2D TMDs and superior light absorbing 0D perovskite nanocrsytals, through proper interface engineering, would open up new pathways for novel optoelectronic functionalities and energy-harvesting applications.

## Acknowledgement:

SKR acknowledges the support of Chair Professor Fellowship of the Indian National Academy of Engineering (INAE).

## Conflicts of interest

There are no conflicts to declare.

# Supplemental Material

# Charge transfer mediated giant photo-amplification in air-stable α-CsPbI$_3$ nanocrystals decorated 2D-WS$_2$ photo-FET with asymmetric contacts

Shreyasi Das[1], Arup Ghorai[1,2], Sourabh Pal[3], Somnath Mahato[1], Soumen Das[4], Samit K. Ray[5] *

[1]*School of Nano Science and Technology, IIT Kharagpur, Kharagpur, West Bengal, India, 721302*
[2]*Department of Materials Science and Engineering, Pohang University of Science and Technology, Pohang 790-784, Korea*
[3]*Advanced Technology Development Centre, IIT Kharagpur, Kharagpur, West Bengal, India, 721302*
[4]*School of Medical Science and Technology, IIT Kharagpur, Kharagpur, West Bengal, India, 721302*
[5]*Department of Physics, IIT Kharagpur, Kharagpur, West Bengal, India, 721302*
Email : physkr@phy.iitkgp.ac.in

## S1: Tauc plot of CsPbI$_3$ NCs

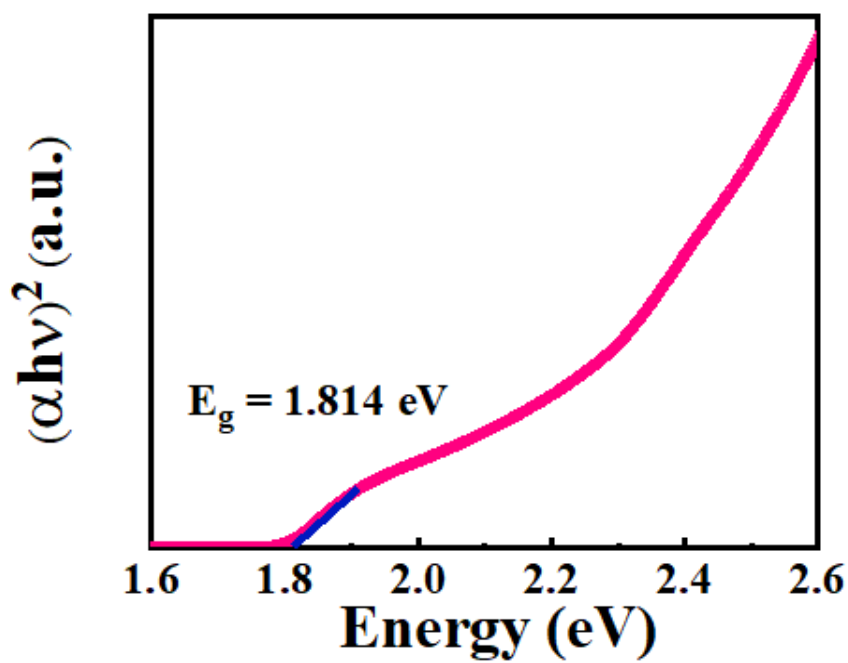

Fig. S1. Tauc plot of as synthesised α-phase CsPbI$_3$ NCs

## S2: Photoluminescence spectra of WS$_2$, CsPbI$_3$ and CsPbI$_3$/WS$_2$ hybrid

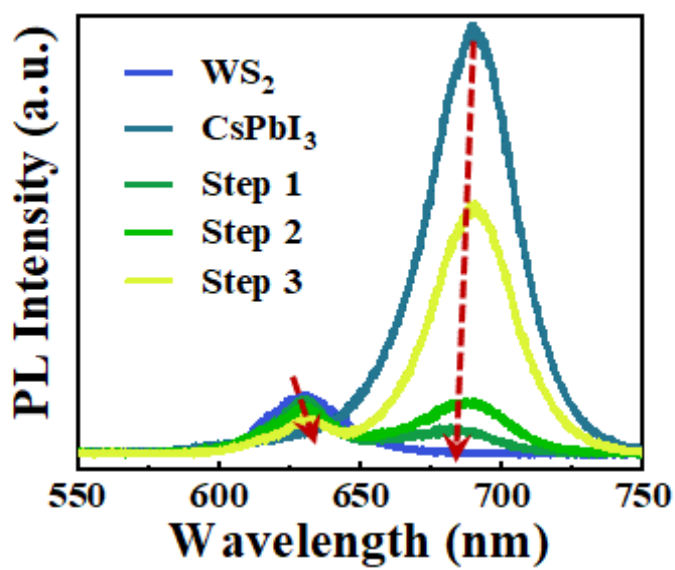

Fig. S2. Comparative PL spectrum of bare ML WS$_2$ flake, CsPbI$_3$ NCs and their heterostructures where three different concentrations of CsPbI$_3$ NCs (different steps of spin coating) incorporated on WS$_2$ flakes. After formation of heterostructures, the PL intensity of both bare ML WS$_2$ as well as CsPbI$_3$ NCs get reduced.

## S3: PL integrated intensity ratio vs coating step

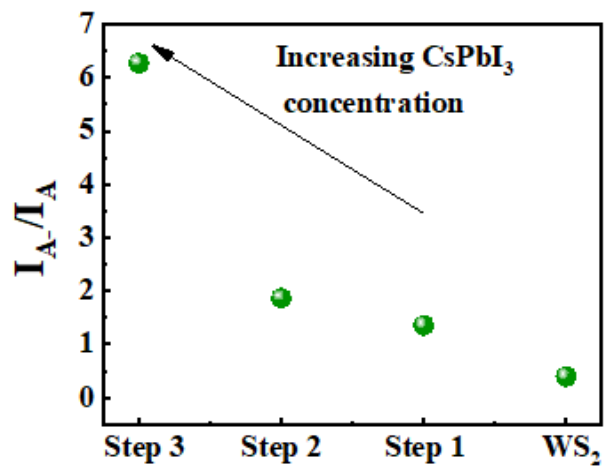

Fig. S3. Variation of PL integrated intensity ratio of $WS_2$ trion peak ($A^-$) to excitonic peak (A) with increasing concentration of $CsPbI_3$ decoration (increasing spin coating step) on $WS_2$ flakes.

## S4: Thickness of the exfoliated flake analysis using AFM

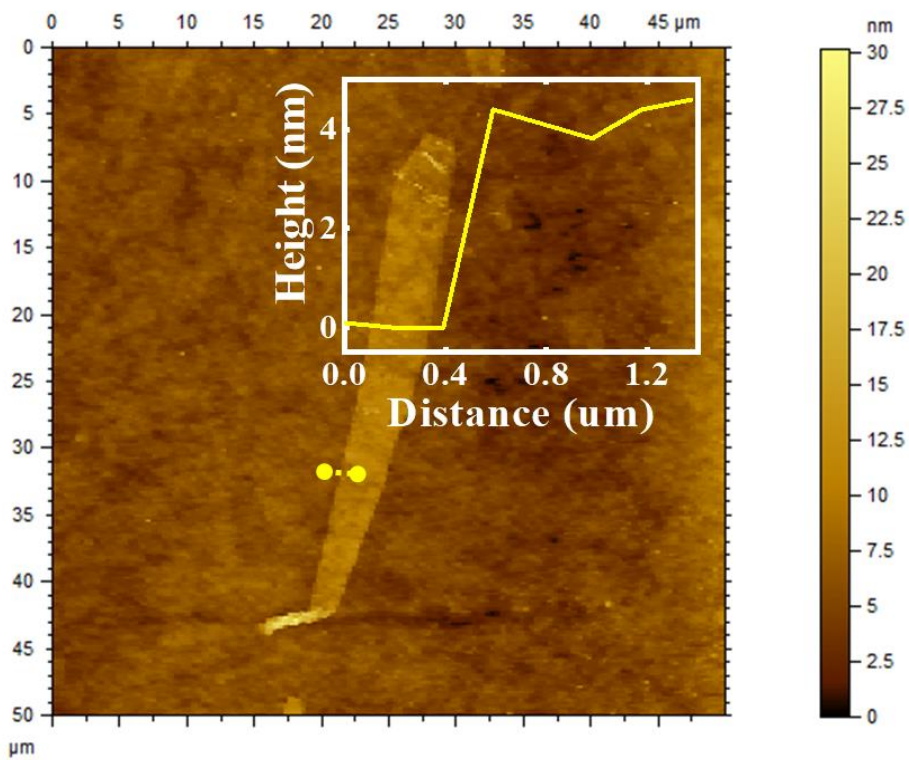

Fig. S4. Atomic force microscopy image of the few layer $WS_2$ flakes. Inset shows the height profile along the yellow dashed line confirming the thickness of the flakes around 4.5 nm.

## S5: Energy band structures of WS₂ at the contacts

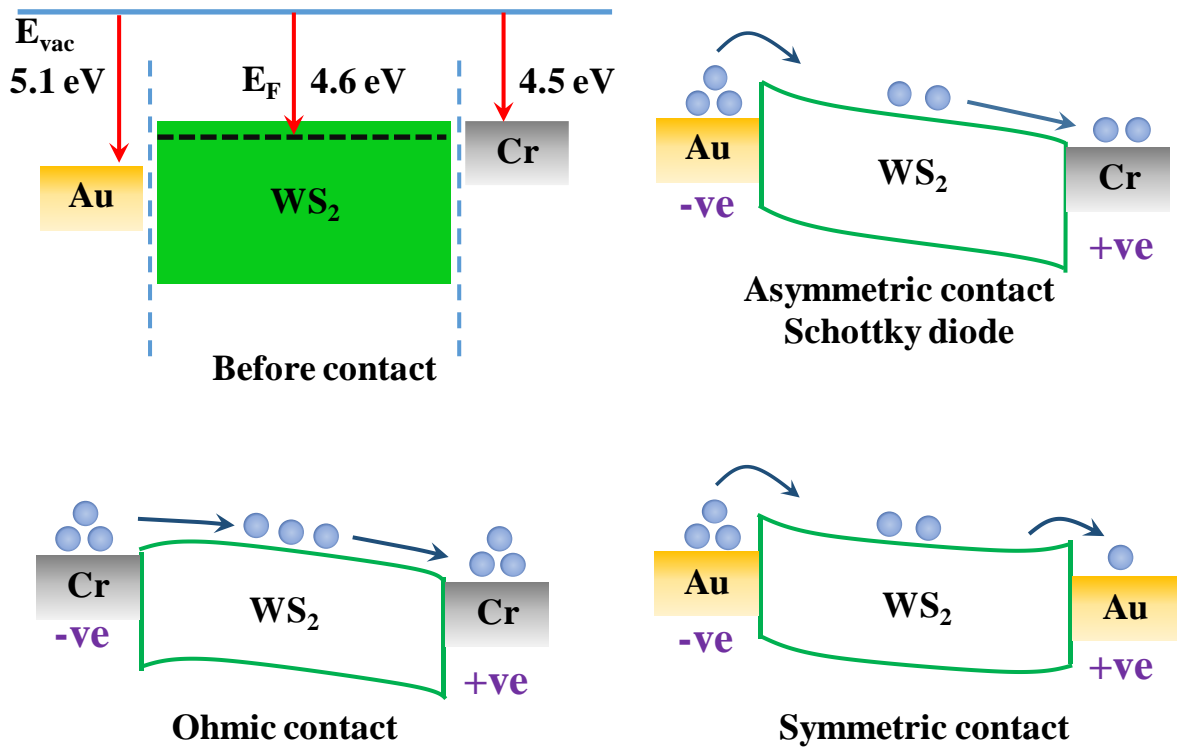

Fig. S5. The corresponding energy band structures with different combination of metal electrodes before contact and after contact condition under applied reverse bias.

## S6: Photo to dark current ratio

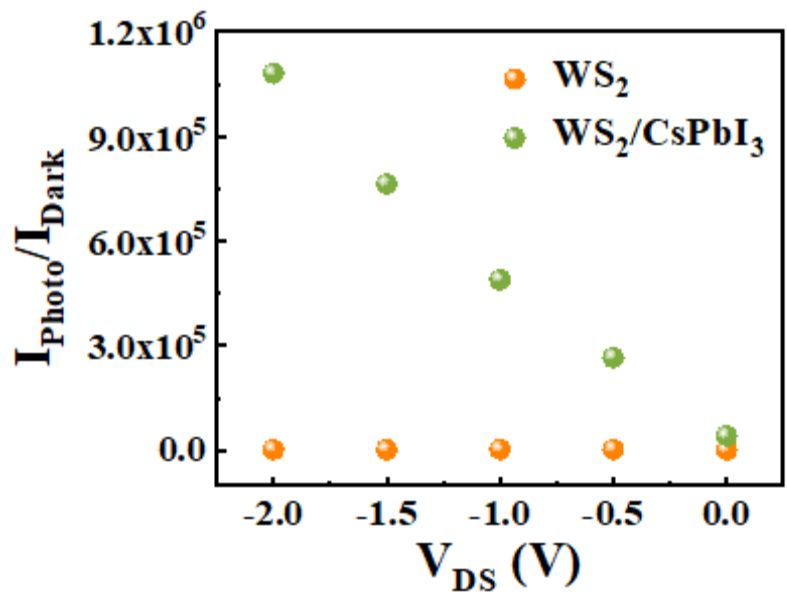

Fig. S6. Photo to dark current ratio for control device (WS₂ FET) and MvWH photo-FET with varying drain to source voltage.

## S7: Spectral responsivity of the control WS₂ FET device

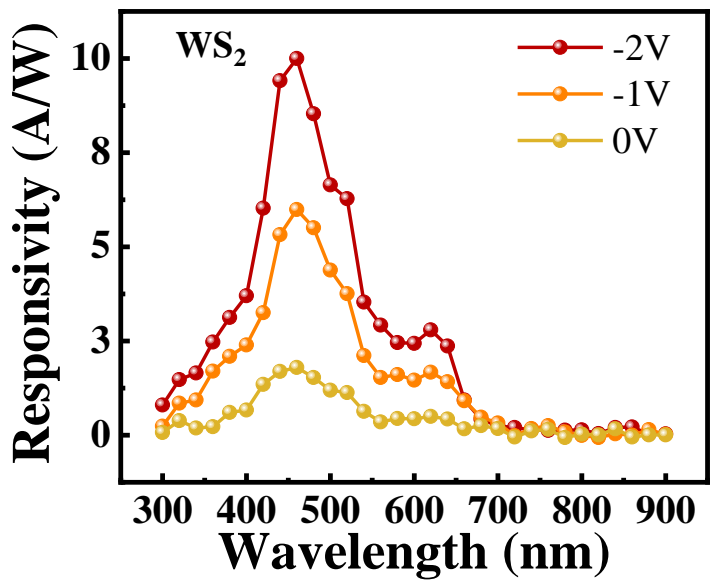

Fig. S7. Spectral responsivity curves of control WS$_2$ FET device at V$_{GS}$ = 0V with increasing reverse V$_{DS}$ from 0V to -2V

## S8: Transient response of the control WS₂ FET device

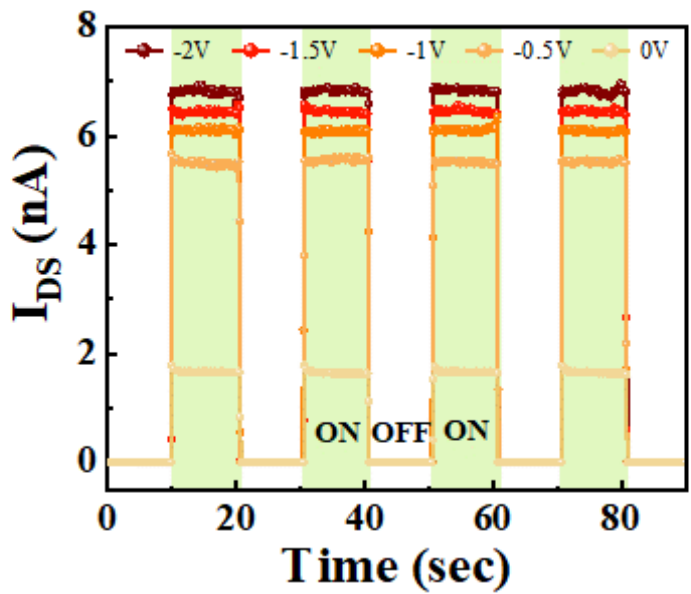

Fig. S8. Transient response of the control WS$_2$ FET device under illumination of a 514 nm laser at different applied V$_{DS}$.